\documentclass[aps,prb,amsmath,amssymb,twocolumn,floatfix]{revtex4-2}
\usepackage{graphicx}
\usepackage{dcolumn}
\usepackage{bm}
\usepackage{epstopdf}

\begin{document}
\title{Coexistence of Kondo Coherence and Localized Magnetic Moments in the Normal State of Molten Salt-Flux Grown UTe$_2$}

\author{N.~Azari,$^1$ M.~Yakovlev,$^1$ S. R.~Dunsiger,$^{1,2}$ O. P.~Uzoh,$^1$ E.~Mun,$^1$ B. M.~Huddart,$^3$ S. J. Blundell,$^3$ 
M. M.~Bordelon,$^4$ S. M.~Thomas,$^4$ J. D. Thompson,$^4$ P. F. S.~Rosa,$^4$ and J. E.~Sonier$^1$}

\affiliation{$^1$Department of Physics, Simon Fraser University, Burnaby, British Columbia V5A 1S6, Canada \\
$^2$Centre for Molecular and Materials Science, TRIUMF, Vancouver, British Columbia V6T 2A3, Canada \\
$^3$Clarendon Laboratory, Department of Physics, University of Oxford, Oxford OX1 3PU, United Kingdom \\
$^4$Los Alamos National Laboratory, Los Alamos, New Mexico 87545, USA}  

\date{\today}
\begin{abstract}
The development of Kondo lattice coherence in UTe$_2$ leads to the formation of a heavy Fermi liquid state from which superconductivity emerges at lower
temperature. In Kondo lattice systems, the nuclear magnetic resonance (NMR) and muon ($\mu^+$) Knight shift have proven to be particularly sensitive to 
the properties of the developing heavy-electron fluid. Here we report $\mu^+$ Knight shift measurements 
on high-quality UTe$_2$ single crystals grown by a molten salt-flux method. Together with previous data from a
single crystal grown by a chemical-vapor transport method, our results show the contribution of the heavy-electron liquid to
the $\mu^+$ Knight shift increases below a crossover temperature $T^* \! \sim \! 30$~K in accord with a universal scaling function of $T/T^*$
for heavy-fermion materials. An observed departure from this universal scaling below a temperature $T_{\rm r} \! \sim \! 12$~K at certain $\mu^+$
stopping sites signifies a reversal of the Kondo hybridization and a relocalization of U $5f$ moments with an antiferromagnetic coupling.
The preservation of universal scaling at a different $\mu^+$ site demonstrates a coexistence of itinerant and localized $5f$electron states 
preceding the superconducting phase transition.   
\end{abstract}
\maketitle

\section{Introduction}
After nearly five years of extensive investigation, UTe$_2$ remains a strong candidate for a rare odd-parity superconductor.
At present, the identity of the superconducting pairing state and the source of the effective attractive interaction responsible for electron pairing are unsettled issues.
In general, the normal state is key to establishing the origin of the superconducting state that emerges at lower temperature.
In UTe$_2$ the normal state immediately above the superconducting transition temperature ($T_c$) exhibits a $T^2$ resistivity and $T$ linear electronic specific heat 
with values of the corresponding coefficients characteristic of a moderate heavy Fermi liquid \cite{Ran:19,Eo:22}. Further evidence of a heavy Fermi-liquid state 
is a constant value of the $^{125}$Te NMR spin-lattice relaxation rate divided by temperature ($1/T_1T$) immediately above $T_c$ 
for applied magnetic fields ${\bf H} \! \parallel \! {\bf b}$ and ${\bf H} \! \parallel \! {\bf c}$ \cite{Tokunaga:19,Ambika:22,Kinjo:22}.
The heavy nature of the itinerant electrons is ascribed to the Kondo effect, where the conduction electron sea gradually screens and hybridizes with 
the lattice of U 5$f$ local moments, with the screening becoming coherent through the lattice below a characteristic temperature $T^*$.

In Kondo latttice systems such as UTe$_2$, there is an expected linear scaling of the Knight shift ($K$) measured by NMR or muon spin rotation ($\mu$SR)  with the 
bulk magnetic susceptibility ($\chi$) for temperatures $T \! > \! T^*$, wherein $\chi$ is dominated by the local moments. However, in many heavy-electron 
materials a breakdown in this scaling is observed below $T^*$ (commonly referred to as a Knight shift or $K$-$\chi$ anomaly) 
that is well described within the framework of a phenomenological two-fluid model \cite{Yang:08,Yang:12}. In the two-fluid description of the Kondo lattice, a
fluid of heavy electrons coexisting with partially screened local moments gradually develops below $T^*$. The loss of scaling arises because of the 
sensitivity of $K$ to the susceptibility of the heavy-electron fluid ($\chi_{\rm HF}$), which exhibits a temperature dependence that is different from the susceptibility 
of the remaining localized moments \cite{Shirer:12}. Yet a significant deviation of the NMR Knight shift from $\chi(T)$ is not observed in UTe$_2$ below $T^*$
\cite{Tokunaga:19,Ambika:22}. By contrast, a clear breakdown in linear scaling of the $\mu^+$ Knight shift ($K_\mu$) with $\chi$ has been observed
in UTe$_2$ below $T^* \! \sim \! 30$~K \cite{Azari:23a}, indicating an enhanced sensitivity of the positive muon to $\chi_{\rm HF}$. Furthermore, below a temperature
$T_{\rm r} \! \sim \! 12$~K there is a gradual return to linear scaling prior to the onset of superconductivity at $T_c$, indicating a relocalization of the local moments.
Nevertheless, a shortcoming of this previous study of $K_{\mu}$ in UTe$_2$ is that it was performed on a large single crystal grown by a chemical vapour transport (CVT) method. 
CVT-grown single crystals exhibit significant differences in $T_c$ and normal state properties due to variations in uranium deficiency \cite{Haga:22}
or atomic displacement parameters \cite{Rosa:22,Weiland:22}. Moreover, they have been shown to contain magnetic clusters \cite{Sundar:23}. 
Consequently, it is unclear whether relocalization is an intrinsic physical property of UTe$_2$. 

Here we report on a $\mu^+$ Knight shift study of higher quality UTe$_2$ single crystals grown by a molten-salt flux (MSF) method. Single crystals grown in this
way are characterized by a high residual resistivity ratio, lower residual specific heat, and the presence of quantum oscillations \cite{Sakai:22,Aoki:22}.

\section{Methods}
Transverse-field (TF) $\mu$SR measurements were performed on the same MSF-grown single crystals ($T_c \! = \! 1.96$~K) investigated in a
recent zero-field (ZF) $\mu$SR study \cite{Azari:23b}, which found no evidence for magnetic clusters and did not detect electronic moments fluctuating slow 
enough to be observable in the $\mu$SR time window. Using the same experimental setup as in Ref.~\cite{Azari:23a}, 
a mosaic of five of the MSF-grown single crystals was attached to a pure silver (Ag) backing plate and TF-$\mu$SR spectra recorded with 
the external magnetic field applied parallel to the crystallographic $c$ axis (${\bf H} \! \parallel \! {\bf c}$). 
Further details of the experimental apparatus and procedure are described in the Supplemental Material \cite{SM}.

The recorded TF-$\mu$SR asymmetry spectrum is a measure of the time evolution of the muon spin polarization $P_\mu(t)$, which was fitted to the function
\begin{equation}
A(t) \! = \! a_0 P_\mu(t) \! = \! \sum_{i = 1}^{n} a_i e^{-\sigma_i^2 t^2} \cos(2 \pi \nu_i \! + \! \phi_i) \, ,
\label{Asymmetry}
\end{equation}
where $a_i$, $\sigma_i$, $\nu_i$ and $\phi_i$ are the amplitudes, depolarization rates, precession frequencies and phase angles of the $n$ components, respectively.
The $\mu^+$ Knight shift for each component is calculated from the precession frequency as follows
\begin{equation}
K_i \! = \! \frac{\nu_i \! - \! \nu_{\rm ext}}{\nu_{\rm ext}} \! + \! 4 \pi \left( \frac{1}{3} - N \right) \rho_{\rm mol} \chi_c \, ,
\label{KnightShift}
\end{equation}
where $\nu_{\rm ext} \! = \! (\gamma_\mu/2 \pi) B_{\rm ext}$ ($\gamma_\mu/2 \pi \! = \! 135.54$~MHz/T) is the muon precession frequency in the
external magnetic field.
The second term in Eq.~(\ref{KnightShift}) is a correction for bulk demagnetization and Lorentz fields, where $N$ is the demagnetization factor, 
$\rho_{\rm mol} \! = \! 0.0186$~mol/cm$^3$ is the molar density of UTe$_2$ and $\chi_c$ is the bulk molar susceptibility for ${\bf H} \! \parallel \! {\bf c}$.
The $\mu^+$ Knight shift is caused by the effective dipolar field at the $\mu^+$ site originating from localized U 5$f$ moments induced by the applied field
and contact hyperfine fields at the $\mu^+$ site from spin polarization of the conduction electrons directly by the applied field and indirectly via a
Ruderman-Kittel-Kasuya-Yosida (RKKY) interaction with the field-induced moments on the localized U 5$f$ electrons. 
\begin{figure}
\centering
\includegraphics[width=\columnwidth]{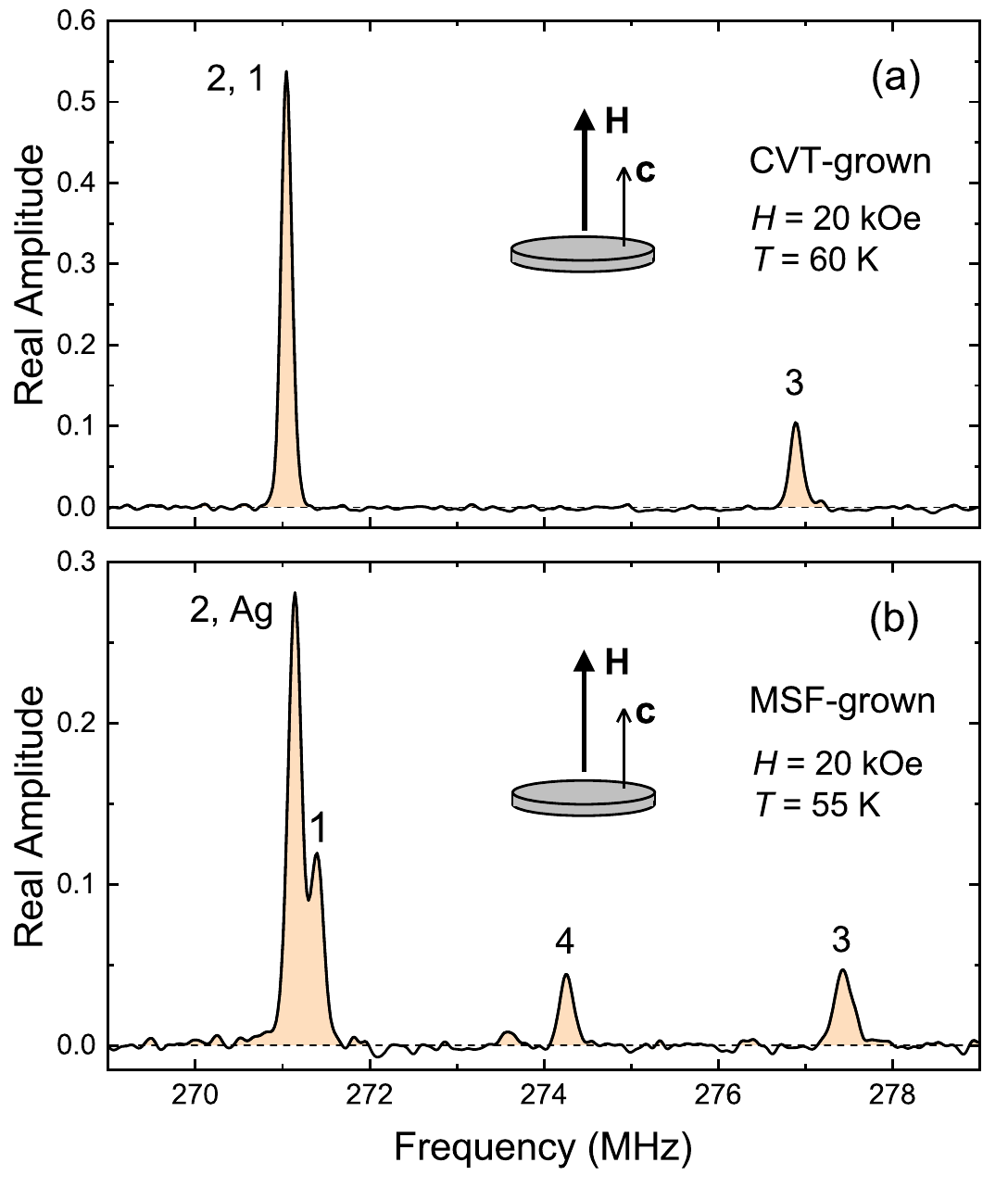}
\caption{Fourier transforms of TF-$\mu$SR asymmetry spectra recorded for (a) CVT-grown \cite{Azari:23a} and (b) MSF-grown single crystals 
in a magnetic field of $H \! = \! 20$~kOe applied parallel to the $c$ axis. In (b) there is a contribution
to the left peak from the Ag backing plate. The numeric labels are chosen for consistency with Ref.~\cite{Azari:23a}.} 
\label{fig1}
\end{figure}
 
\subsection{Muon Knight shift} 
Figure~\ref{fig1} shows a comparison of the Fourier transforms of a TF-$\mu$SR asymmetry spectrum in the previously studied CVT-grown single crystal and the
MSF-grown single crystal mosaic recorded at a temperature well above $T_c$ and for a field $H \! = \! 20$~kOe applied parallel to the $c$ axis.
Although two peaks are observed in the Fourier transform for the CVT-grown single crystal, fits in the time domain reveal that the left peak is comprised of two
closely spaced frequencies (denoted 1 and 2) \cite{Azari:23a}. In the MSF-grown single crystals these two frequencies are visually split.
Moreover, due to spacing between the single crystals there is a signal contributing to the leftmost peak from muons that stopped in 
the Ag backing plate, which is well resolved in the time domain. There is also an additional frequency [labelled 4 in Fig.~\ref{fig1}(b)] not observed in 
the CVT-grown single crystal. This frequency is also not observed in TF-$\mu$SR spectra recorded for one of the five
MSF-grown single crystals (see Supplemental Material \cite{SM}) and hence is presumably due to a misalignment of one or more of the five single crystals in the mosaic
with the applied field. 
The amplitudes of the different components of the TF-$\mu$SR asymmetry spectrum are shown in Table~\ref{amplitudes} along with
those for the previously studied CVT-grown single crystal and for a few measurements of one of the MSF-grown single crystals.
\begin{table}
\caption{The amplitudes of the individual components of Eq.~(\ref{Asymmetry}) for one of the MSF-grown single crystals, the MSF-grown single crystal mosaic, 
and the CVT-grown single crystal of Ref.~\cite{Azari:23a}.}
\begin{ruledtabular}
\begin{tabular}{cccc}
{\bf  } & MSF-grown & MSF-grown & CVT-grown \\
{\bf  } & single crystal & mosaic & single crystal \\ \hline
  $a_1$ &34.1(2)~\%&31(1)~\%&27(2)~\% \\
  $a_2$ &54.7(3)~\%&44(2)~\%&55(2)~\% \\
  $a_3$ &11.1(1)~\%&14(1)~\%&18(1)~\% \\
  $a_4$ &- &11(1)~\%&- \\
\end{tabular}
\end{ruledtabular}
\label{amplitudes}
\end{table}
\begin{figure*}
\centering
\includegraphics[width=\textwidth]{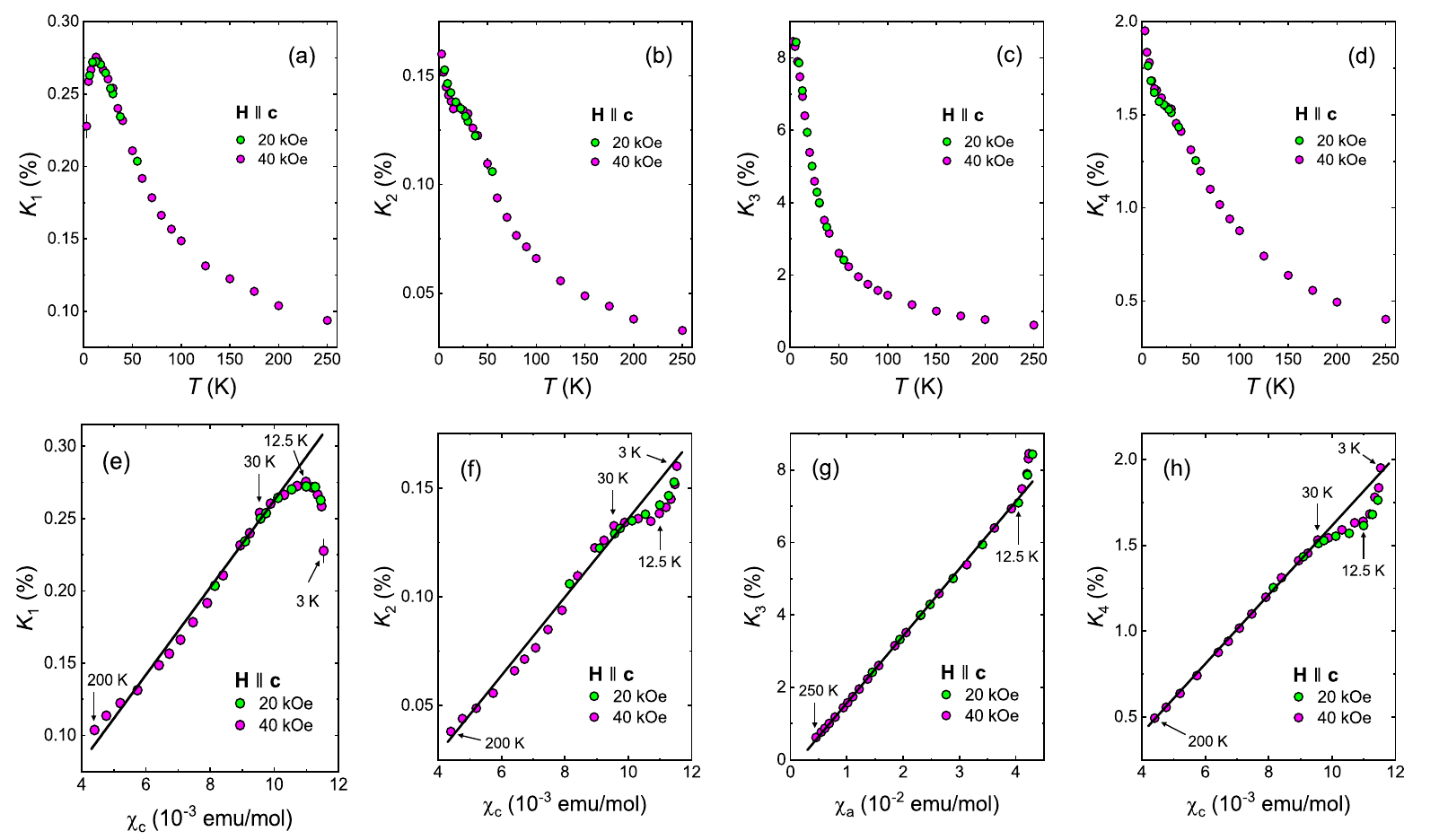}
\caption{(a)-(d) Temperature dependence of the four $\mu^+$ Knight shift components in the mosaic of MSF-grown UTe$_2$ single crystals for a magnetic field applied parallel to the $c$ axis.
(e)-(h) The four $\mu^+$ Knight shift components for ${\bf H} \! \parallel \! {\bf c}$ versus the bulk magnetic susceptibility. (e) $K_1$, (f) $K_2$ and (h) $K_4$ are plotted versus 
the bulk susceptibility for ${\bf H} \! \parallel \! {\bf c}$, whereas (g) $K_3$ is plotted versus the bulk susceptibility for ${\bf H} \! \parallel \! {\bf a}$.}
\label{fig2}
\end{figure*}

While initial density functional theory (DFT) calculations identified a single muon stopping site in UTe$_2$ based solely 
on the minima of the electrostatic potential \cite{Sundar:23}, the different components of the TF-$\mu$SR spectrum indicate three distinct muon sites.
Note that $a_2 \! + \! a_4$ for the MSF-grown single crystal mosaic is equivalent to $a_2$ for the other two samples in Table~\ref{amplitudes},
suggesting that the fourth component originates from the same muon stopping site as that associated with $a_2$ and that this site is most 
influenced by a misalignment with respect to the applied field.
DFT calculations taking into account distortions of the local environment by the muon identify five candidate muon sites
(see Supplemental Material \cite{SM}). Two of these are of significantly lower energy and lie between nearest-neighbor U atoms within the two-leg ladder 
substructure formed by the U atoms. One is a stable site midway between U atoms forming the ladder rungs along the $c$ axis, while the other 
is midway between nearest-neighbor U atoms in the ladder legs parallel to the $a$ axis.
Inelastic neutron scattering (INS) \cite{Knafo:21} and NMR \cite{Tokunaga:19,Ambika:22}
measurements on CVT-grown samples indicate ferromagnetic (FM) coupling of the U atoms within the ladders and antiferromagnetic (AFM) coupling between the U-ladders.
Muon sites within the U-ladders may explain why an initial ZF-$\mu$SR study of CVT-grown UTe$_2$ found evidence for FM fluctuations \cite{Sundar:19}.
The muon may also stop in any of the three higher-energy sites found by DFT. However, the population ratio of the three muon sites
inferred by TF-$\mu$SR ($a_1 \! : \! a_2 \! : \! a_3$) does not match any combination of the DFT-computed candidate sites,
and is incompatible with the multiplicities of positions with the $Immm$ space group, suggesting that the muon is not equally likely to stop in any given muon site.
\begin{figure*}
\centering
\includegraphics[width=18cm]{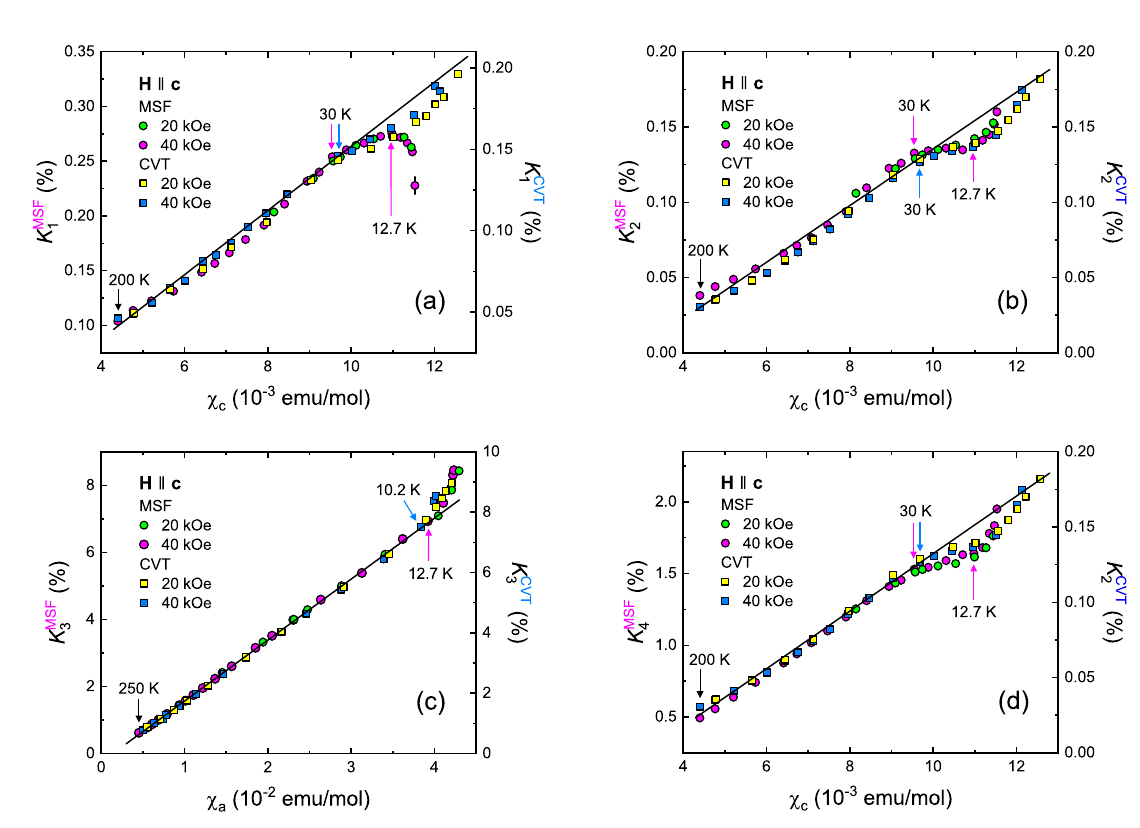}
\caption{Comparison of the $\mu^+$ Knight shift components for the MSF-grown UTe$_2$ single-crystal mosaic with those of the CVT-grown single crystal 
of Ref.~\cite{Azari:23a}. (a) $K_1$ versus $\chi_c$, (b) $K_2$ versus $\chi_c$, and (c) $K_3$ versus $\chi_a$ for both samples with temperature as an
implicit parameter. (d) A comparison of $K_4$ versus $\chi_c$ for the MSF-grown sample with $K_2$ versus $\chi_c$ for the CVT-grown sample. The solid 
line through the data points in each panel is a guide to the eye. Note the vertical scales in (a), (c) and (d) are different for the two samples.}
\label{fig3}
\end{figure*}

Figure~\ref{fig2}(a)-(d) shows the temperature dependence of the four components of the $\mu^+$ Knight shift in the MSF-grown UTe$_2$ sample.
As shown in Fig.~\ref{fig2}(e)-(h), three of the components exhibit linear scaling with the bulk magnetic susceptibility for ${\bf H} \! \parallel \! {\bf c}$ ($\chi_c$) above 
$T \! \sim \! 30$~K, whereas $K_3$ instead tracks the bulk susceptibility for ${\bf H} \! \parallel \! {\bf a}$ ($\chi_a$) down to $T \! \sim \! 12$~K. 
This strange behavior of $K_3$ was also observed in the CVT-grown sample [see Fig.~\ref{fig3}(c)]. While we previously speculated that this originates from muons 
stopping in magnetic clusters, we now know that these clusters are not present in the MSF-grown single crystals.
The large size of $K_3$ suggests this component originates from muons experiencing large hyperfine fields at sites close to the U atoms.
Calculations based on a crystal electric field (CEF) Hamiltonian calculated by considering the Coulomb repulsion between the central U$^{4+}$ ion and the 
surrounding ligands, which are treated as point charges, show that in principle the local susceptibility sensed by a muon
near U may closely resemble $\chi_a$ due to muon-induced local lattice distortions and a greater effect of the
muon charge on the CEF levels of U (see Supplemental Material \cite{SM}).
Unfortunately, the CEF parameters obtained from this highly approximate point-charge model
bear little resemblance to those actually required to describe the temperature dependence of $\chi_a$, $\chi_b$ and $\chi_c$ \cite{Rosa:22}.

The data presented in Fig.~\ref{fig2} supports the earlier assertion that $K_4$ is associated with the same muon stopping site as $K_2$,
but appears to be somewhat influenced by a misalignment with the applied field that introduces some component of $\chi_a$. 
Since $K_3 \! \gg \! K_2$, this explains the larger muon Knight shift for $K_4$, but a broader linewidth is also expected. 
Indeed, the depolarization rate $\sigma_4$ is significantly greater than $\sigma_2$ and closer to $\sigma_3$ (see Supplemental Material \cite{SM}).  

\subsection{Comparison between MSF and CVT-grown single crystals}
The behaviors of $K_2$ and $K_4$ are similar to that of $K_1$ and $K_2$ in the CVT-grown sample (see Fig.~\ref{fig3}) in displaying a breakdown
in linear scaling with $\chi_c$ below the temperature $T^* \! \sim \! 30$~K and a gradual return to tracking $\chi_c(T)$ as the temperature is 
lowered below $T_{\rm r} \! \sim \! 12$~K. 
The loss of scaling below $T^* \! \sim \! 30$~K was previously attributed to the
change that occurs in the contact hyperfine field at the muon site due to the interaction of the muon spin with the unpaired spin density of the conduction 
electrons upon the onset of Kondo coherence \cite{Azari:23a}. The sensitivity of the muon Knight shift to the onset of Kondo coherence
has also been demonstrated through the observation of a Knight shift anomaly in the $4f$-electron Kondo insulator SmB$_6$ \cite{Akintola:18}.
In contrast, $K_1$ in the MSF-grown sample does not return to tracking $\chi_c$ as the temperature is lowered below $T_{\rm r}$. This behavior and the 
peak in the plot of $K_1$ versus $T$ [see Fig.~\ref{fig2}(a)] were not observed in the CVT-grown single crystal \cite{Azari:23a}.
We suggest this is due to the presence of the magnetic clusters in the CVT-grown sample, which gives rise to a low-field upturn in the
bulk magnetic susceptibility below $T \! \sim \! 10$~K \cite{Azari:23a} that is not observed in the MSF-grown single crystals \cite{Azari:23b}.
As we have previously argued, the magnetic clusters presumably are the result of local disorder/defect induced disruptions of long-range 
correlations along the U-ladders and this may in turn disrupt the Kondo coherence sensed by the muon at the site associated with $K_1$.
As shown in Fig.~\ref{fig3}(a), it is below $T \! \sim \! 12$~K where the temperature dependence of $K_1$ in the two samples deviates.

\subsection{Heavy-electron contribution to the muon Knight shift}
\begin{figure}
\centering
\includegraphics[width=\columnwidth]{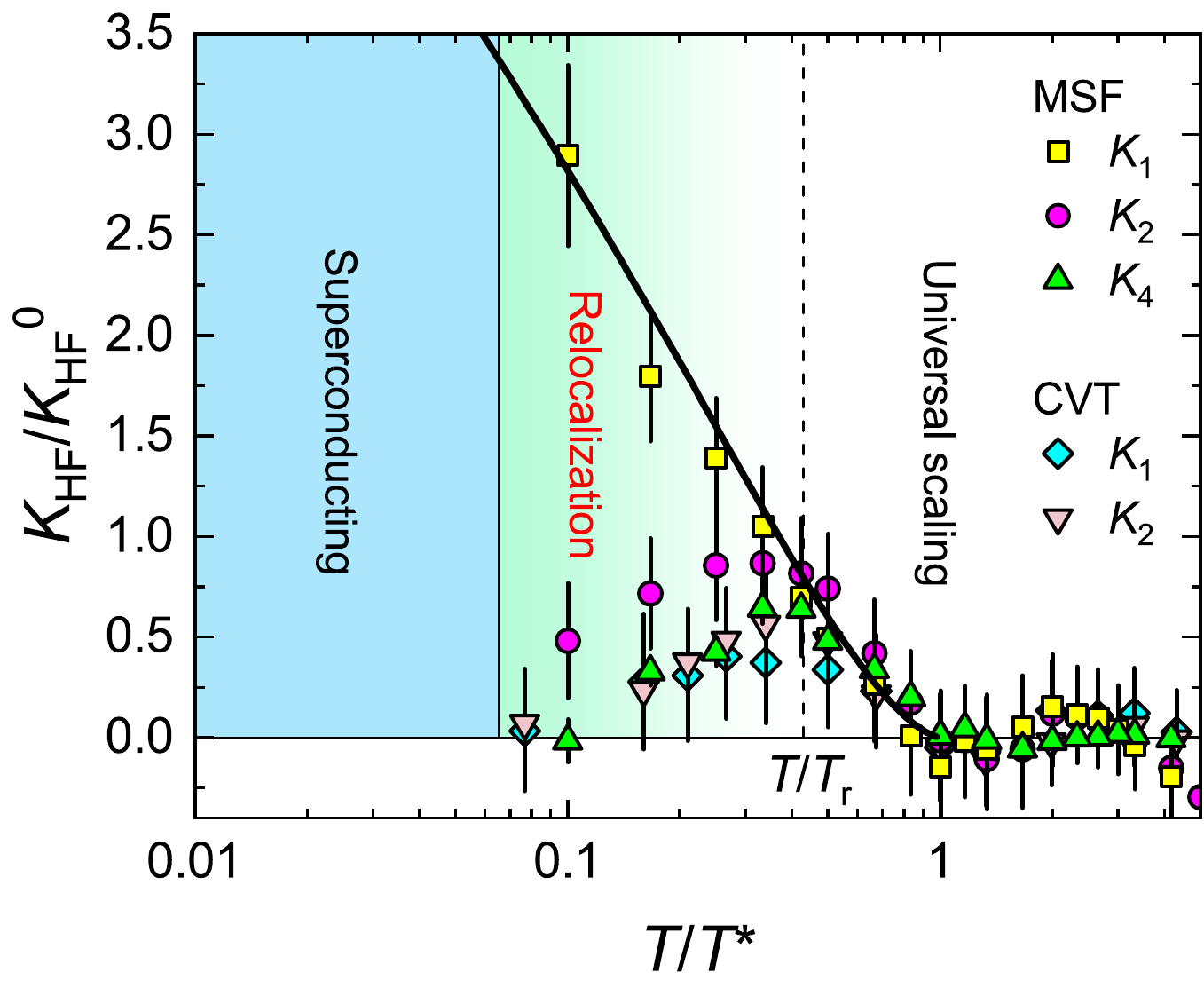}
\caption{Plot of the normalized heavy-electron fluid contribution to the individual components of the $\mu^+$ Knight shift in the MSF-grown single 
crystal mosaic and the CVT-grown single crystal of Ref.~\cite{Azari:23a} assuming $T^* \! = \! 30$~K. All data are for $H \! = \! 40$~kOe applied parallel to the $c$ axis.
The solid curve is Eq.~(\ref{KHF}). The label $T/T_{\rm r}$ corresponds to $T_{\rm r} \! = \! 12$~K.}
\label{fig4}
\end{figure}

In a wide range of materials the heavy-electron contribution to the NMR or $\mu^+$ Knight shift anomaly obeys a universal scaling behavior 
predicted by the two-fluid model \cite{Yang:08}
\begin{equation}
K_{\rm HF}(T) \! = \! K_{\rm HF}^{0} (1 - T/T^*)^{3/2}[1 + \log(T/T^*)] \, ,
\label{KHF}
\end{equation}
where $K_{\rm HF}^{0}$ and $T^*$ are material dependent quantities.
The temperature dependence of the emergent heavy-electron fluid for the individual components of the $\mu^+$ Knight shift in UTe$_2$ is determined 
by subtracting a linear extrapolation of the data for $K_i(T)$ versus $\chi_c(T)$ above $T^*$ in the range 30~K~$< \! T \! <  150$~K. 
Figure~\ref{fig4} shows a plot of $K_{\rm HF}/K_{\rm HF}^{0}$ versus $T/T^*$ for the $\mu^+$ Knight shift components 
of the MSF-grown single crystal mosaic and the CVT-grown single crystal of Ref.~\cite{Azari:23a} assuming $T^* \! = \! 30$~K. 
The plot excludes $K_3$, which accounts for 14.4~\% and 18~\% of the TF-$\mu$SR signal for the MSF-grown and CVT-grown samples, respectively.  
While the heavy-electron contribution to $K_1$ of the MSF-grown sample follows Eq.~(\ref{KHF}) down to the onset of superconductivity, 
the contribution of $K_{\rm HF}$ to $K_2$ and $K_4$ deviates from the scaling equation below the temperature $T_{\rm r} \! \sim \! 12$~K, as previously observed in 
CVT-grown UTe$_2$. Anomalies have been observed in various physical quantities near $T \! \simeq \! 12$~K, including a minimum in the electronic contribution 
to the $c$-axis thermal expansion, and broad peaks in the electronic specific heat, temperature derivative of the $a$-axis electrical resistivity, and 
the $c$-axis resistivity \cite{Willa:21,Eo:22,Thebault:22}. 

The downturn in $K_{\rm HF}(T)$ below Eq.~(\ref{KHF}) that commences at $T_{\rm r} \! < \! T^*$ signifies a reversal of the
hybridization of the conduction electrons with the local U $5f$-electron moments and a gradual relocalization of the local moments as a precursor
to a magnetically-ordered ground state \cite{Shirer:12}. To date this behavior has only been observed in materials that order antiferromagnetically. 
In CeRhIn$_5$ \cite{Shirer:12}
and CePt$_2$In$_7$ \cite{apRoberts:11}, $K_{\rm HF}$ determined by NMR deviates from Eq.~(\ref{KHF}) below a temperature $T_{\rm r}$ at which
AFM correlations begin to develop between the partially screened Ce $4f$ local moments prior to the onset of an AFM phase at lower temperature.
By analogy it appears that short-range AFM interactions between relocalizing U $5f$ local moments in UTe$_2$ commence at $T_{\rm r} \! \sim \! 12$~K.
Near this temperature the AFM fluctuations observed by INS in CVT-grown samples saturate \cite{Knafo:21,Duan:20,Butch:22}. 
It has been suggested that this temperature marks a crossover
from the development of quasi-1D FM fluctuations from noninteracting U-ladders to quasi-2D magnetic fluctuations
driven by AFM interactions between the U-ladders \cite{Thebault:22}. In this scenario, the Knight shift components
indicating relocalization presumably correspond to sites where the $\mu^+$ is sensitive to interladder AFM interactions.

Relocalization of itinerant $f$ electrons leading up to an AFM phase transition has also been observed by angle-resolved photoemission spectroscopy (ARPES)
in the $4f$-electron compounds CePt$_2$In$_7$ \cite{Luo:20} and CeCoGe$_3$ \cite{Li:23}, and recently in the $5f$-electron compound UPd$_2$Al$_3$ \cite{Song:24}. 
In UPd$_2$Al$_3$ the transition to long-range AFM order is followed by a superconducting transition at much lower temperature.
While UTe$_2$ does not undergo a magnetic phase transition \cite{Sundar:19}, an AFM phase is induced by pressure \cite{Thomas:20}, which has
recently been shown to have an incommensurate magnetic wavevector close to that of the AFM fluctuations at ambient pressure \cite{Knafo:24}.
This suggests that the Kondo coherence is suppressed by pressure and UTe$_2$ is close to an AFM instability at ambient pressure. 
Such a scenario has been proposed for the Kondo insulator CeRu$_4$Sn$_6$, wherein relocalization of the Ce $4f$ electrons has been observed by ARPES,
despite no evidence for magnetic order \cite{Wu:23}. 

It is apparent from the behavior of $K_{\rm HF}(T)$ derived from $K_1$ in the MSF-grown UTe$_2$ sample that itinerant heavy electrons from
Kondo hybridization coexist with AFM fluctuations associated with interacting relocalized U $5f$ moments that contribute to other components of the $\mu^+$ Knight shift. This coexistence appears to be a consequence of the dual itinerant and localized nature of the $5f$ electrons. 
Coexisting Kondo hybridization and magnetic order through $5f$-electronic orbital selectivity has been observed in a few U-based compounds.
Scanning tunneling microscopy (STM) and ARPES studies of the AFM materials USb$_2$ and UAs$_2$ show that Kondo hybridization occurs deep 
inside the AFM state, with the antiferromagnetism and itinerant heavy electrons residing on predominantly different $5f$-orbitals that give rise to two
distinct flat bands \cite{Giannakis:19,Chen:19,Ji:22}. Kondo hybridization coexisting with itinerant $5f$-electron ferromagnetism has also been 
observed in an STM study of UGe$_2$ \cite{Giannakis:22}. 

In the case of UTe$_2$, an unsettled issue has been whether the itinerant U$^{3+}$ ($5f^3$) or localized U$^{4+}$ ($5f^2$) state dominates the mixed valence at 
ambient pressure. Recent high-resolution resonant inelastic x-ray scattering measurements provide a fingerprint that the $5f^2$ valence state is the dominant
 multiplet configuration \cite{Christovam:24}. Nonetheless, in a mixed valence metal such as UTe$_2$, the effective valence will likely be $5f^{2+\delta}$, wherein 
extra $5f$ electrons hybridize with conduction electrons. Hence, our simultaneous observation of relocalization and heavy-electron fluid behavior in 
the $\mu^+$ Knight shift appears to arise from having both $5f^2$ local orbitals and additional delocalized $5f$ electrons due to strong hybridization 
with ligand orbitals.
Theoretically it has been shown that the Kondo effect in UTe$_2$ is orbital selective and anisotropic, resulting from hybridization of certain 
U $5f$ orbitals with U $6d$ conduction electrons in the $ab$-plane and hybridization of U $5f$ orbitals with Te $5p$ conduction electrons 
along the $c$ axis at much lower temperature \cite{Kang:22}.
Because of this anisotropy, the change in the contact hyperfine field due to the interaction of the muon spin with the conduction electrons that arises upon formation
of the heavy-electron fluid may be different at different muon sites. In particular, $K_1$ in MSF-grown UTe$_2$ may be attributed 
to a stopping site where the muon senses U $5f$ orbitals that remain delocalized through the Kondo interaction as the temperature is lowered towards $T_c$, 
whereas $K_2$ (and $K_4$) appears to originate from a site where the muon senses U $5f$ orbitals that are driven through a relocalization transition.

\section{Conclusions}
In summary, we have shown through the behavior of the $\mu^+$ Knight shift in high quality single crystals that relocalization of the U $5f$ electrons
prior to the onset of superconductivity is an intrinsic physical property of UTe$_2$. Our findings reveal a dual localized and
itinerant nature of the $5f$ electrons at ambient pressure that manifests as a coexistence of Kondo hybridization and AFM interactions 
between localized moments.  

\begin{acknowledgments}
J.E.S., S.R.D. and E. Mun acknowledge support from the Natural Sciences and Engineering Research Council of Canada. 
Work at Los Alamos National Laboratory by S.M.T., J.D.T. and P.F.S.R. was supported by the U.S. Department of Energy, Office of Basic Energy Sciences,
Division of Materials Science and Engineering. Work at the University of Oxford by B.M.H. and S.J.B. was funded by UK Research and Innovation (UKRI) 
under the UK government’s Horizon Europe guarantee funding (Grant No. EP/X025861/1).
\end{acknowledgments}

\end{document}